\documentclass[journal]{IEEEtran}
\usepackage[utf8]{inputenc}
\usepackage[T1]{fontenc}
\usepackage{scalerel}
\usepackage{amsmath,color}
\usepackage{amsthm,amsfonts}
\usepackage{latexsym}
\usepackage{graphicx}
\usepackage{url}
\usepackage{tikz}

\usepackage{pgfplots}
\usepackage[colorlinks]{hyperref}
\usepackage{xcolor}

\hypersetup{
colorlinks,
    linkcolor={red!50!black},
    citecolor={blue!50!black},
    urlcolor={blue!20!black}
}
\usepackage[utf8]{inputenc}
\usepackage{todonotes}
\presetkeys{todonotes}{inline}{}
\usepackage{subcaption}
\usepackage{bm}

\colorlet{colorMG}{blue}
\colorlet{colorDS}{green!75!black}
\colorlet{colorCE}{yellow}
\colorlet{colorLJ}{orange}
\colorlet{colorAO}{red}
\colorlet{colorSN}{red}
\colorlet{colorYI}{teal!75}
\colorlet{colorMN}{red!50}
\colorlet{colorJL}{pink}
\colorlet{colorAE}{cyan}

\usepackage[normalem]{ulem}

\graphicspath{{figures/}}
\newcommand{\ie}{\textit{i.e.}\/, }
\newcommand{\eg}{\textit{e.g.}\/, }

\newcommand{\toh}{\hat{\to}}

\providecommand*{\mrm}[1]{\mathrm{#1}}
\providecommand*{\unit}[1]{\ensuremath{\mrm{\,#1}}}
\renewcommand{\vec}[1]{{\boldsymbol#1}}
\providecommand*{\unit}[1]{\ensuremath{\mrm{\,#1}}}

\providecommand*{\iu}{\ensuremath{\mrm{i}}}

\providecommand*{\cu}{\ensuremath{c}}
\renewcommand{\Im}{\operatorname{Im}}	
\providecommand*{\diff}{\operatorname{d}\!}
\newcommand{\C}{\mathbb{C}}	

\newcommand{\zvh}{\hat{\vec{z}}}

\usepackage{thmtools}

\begin{document}

\title{Fundamental bounds on transmission through periodically perforated metal screens with experimental validation}
\author{Andrei~Ludvig-Osipov,
Johan~Lundgren,
Casimir~Ehrenborg,
Yevhen~Ivanenko, \\
Andreas~Ericsson, 
Mats~Gustafsson, 
B.L.G.~Jonsson, 
Daniel~Sj{\"o}berg
\thanks{
A.~Ludvig-Osipov and B.L.G.~Jonsson are with the School of Electricl Engineering and Computer Science, KTH Royal Institute of Technology, Stockholm SE-10044, Sweden (e-mail: osipov@kth.se; ljonsson@kth.se).

Johan~Lundgren, Casimir~Ehrenborg, Mats~Gustafsson, and~Daniel~Sj{\"o}berg are with the Department of Electrical and Information Technology, Lund University, Lund SE-22100, Sweden (e-mail: johan.lundgren@eit.lth.se; casimir.ehrenborg@eit.lth.se; mats.gustafsson@eit.lth.se; daniel.sjoberg@eit.lth.se).

Yevhen~Ivanenko is with Department of Physics and Electrical Engineering, Linn\ae us University, V{\"a}xj{\"o} SE-35195, Sweden (e-mail: yevhen.ivanenko@lnu.se).

Andreas~Ericsson was with Lund University, Lund SE-221 00, Sweden. He is now with TICRA, Copenhagen DK-1119, Denmark (e-mail: ae@ticra.com). 
}}

\maketitle

\begin{abstract}
This paper presents a study 
of transmission through arrays of periodic sub-wavelength apertures.
Fundamental limitations for this phenomenon are formulated as a sum rule, relating the transmission coefficient over a bandwidth to the static polarizability. The sum rule is rigorously 
derived for arbitrary periodic apertures in thin screens.
By this sum rule we establish a physical bound on the transmission bandwidth which is verified numerically for a number of aperture array designs.
We utilize the sum rule to design and optimize sub-wavelength frequency selective surfaces with a bandwidth close to the physically attainable.
Finally, we verify the sum rule and simulations by measurements of an array of horseshoe-shaped slots milled in aluminum foil.

\end{abstract}

\section{Introduction}
High transmission through periodically perforated metal screens occurs at certain frequencies associated with various resonance phenomena.
Those can be intrinsic resonances of the apertures, related to their size and geometry, or resonances related to the periodicity.
Some transmission peaks, referred to as extraordinary transmission~\cite{Ebbesen+etal1998,Koerkamp+etal2004,Moreno+etal2001}, have been explained from a theoretical perspective by surface plasmon polaritons at optical frequencies~\cite{Koerkamp+etal2004,Genet+Ebbesen2007}, and by spoof plasmons at radio frequencies~\cite{Pendry+etal2004}.
From the practical perspective it is desirable to know what is the maximum attainable transmission bandwidth, regardless of the nature of the transmission.
Here, we formulate a fundamental bound on the transmission bandwidth in the form of a sum rule.

Applications of such structures include spatially tunable filters, near-field imaging and modulators as well as negative refractive index metamaterials~\cite{Meca+etal2009}.
In frequency selective surface (FSS) design subwavelength apertures are commonly used~\cite{Munk2000}.
A proposed FSS application is slotted infrared-protective metalized windows~\cite{Gustafsson+etal2006}. These would be transparent for cell phone signals, to increase coverage inside of buildings, while serving as a barrier for infrared waves. 
Also, such structures localize high power flow within the apertures~\cite{Beruete+etal2006}. This effect can be additionally increased by designing apertures with narrow slots. This can be used to create nonlinear devices with strong concentration of fields. 
A further application is the Bethe-hole directional coupler~\cite{Bethe1944,Ishii1995}, where a periodic sequence of apertures in a wall joining two waveguides is designed to provide a coupling mechanism in the band of interest. 
Extraordinary transmission (EoT) is an interesting phenomenon, when transmissivity exceeds the predictions of a classical diffraction theory~\cite{Bethe1944,Ishii1995}.
Originally discovered and analyzed in optics~\cite{Ebbesen+etal1998}, it has been observed and discussed also for electromagnetic millimeter waves, see \eg~\cite{Beruete+etal2004,Medina+etal2008}, and in acoustics~\cite{Lu+etal2007}.

A limiting factor of many FSS and EoT applications is that the frequency bands where a high transmission occurs are rather narrow.
Naturally, it is desirable to understand the limitations of this effect; how much bandwidth is achievable and at what frequencies.
The tuning of the transmission bandwidth is an iterative trial-and-error procedure, usually assisted by heuristically reasoned guidelines
{~\cite{Munk2000}}.
To facilitate this procedure, a physical bound in the form of a sum rule is proposed. 
Sum rules have been derived for different types of periodic structures, such as transmission blockage~\cite{Gustafsson+etal2009,Sjoberg+etal20101}, extinction cross section~\cite{Gustafsson+etal2012}, high-impedance surfaces~\cite{Gustafsson+Sjoberg2011}, and antennas~\cite{Doane+etal2013,Jonsson+etal2013}. The transmission cross section sum rule~\cite{Gustafsson2009} produces a bound on the bandwidth transmission cross section product for single apertures. These sum rules are targeted for their particular applications and cannot be used to determine bounds on the perforated screen's transmission bandwidth.

In this paper, we present a derivation of a sum rule for perforated screens, which shows that the total transmission bandwidth is limited from above by the normalized static polarizability of the structure. We validate the sum rule by comparison both with simulated periodic structures, and also with measurements.
We illustrate how the sum rule can be utilized in the design and bandwidth optimization of FSS. 
The sum rule is derived here for structures consisting of an infinitely thin perfect electric conductor (PEC) {screen}. 
We show by simulations that the transmitted power and bandwidth of a generic periodic design is not greatly affected by a finite thickness and conductivity up to certain limits.
This motivates the use of the sum rule in evaluation of real structures as well as in a penalty function in optimization.
The sum rule is valid for all types of transmission peaks for periodically perforated metal screens.

In the presented examples, we consider transmission peaks, associated with resonances of sub-wavelength apertures.
A horseshoe slot aperture was designed and optimized utilizing the sum rule to maximize the transmission bandwidth in the 
lowest frequency
peak.
This design was manufactured in aluminum foil and measured in the frequency range $10\unit{GHz}$ to $20\unit{GHz}$.

The rest of the paper is organized as follows. 
Section~\ref{sec:Theory1} formulates the problem of scattering against periodic screens, and Section~\ref{sec:derivation} gives a derivation of the sum rule for periodic structures.
Numerical examples validating and illustrating the sum rule are presented in Section~\ref{sec:num} along with a demonstration of how the sum rule is used in the design process. Section~\ref{sec:Practice} investigates the applicability of the sum rule for non-ideal structures.
Section~\ref{sec:Measurements} provides the details of the manufacturing process and the measurement setup, and presents the measured transmission coefficient, which is also compared to the theoretical predictions.
Finally, the results of this paper are summarized and discussed in Section~\ref{sec:Conclusions}.

\section{Scattering by periodic perforated screens}\label{sec:Theory1}

\begin{figure}
\centering
   \includegraphics[width=0.6\linewidth]{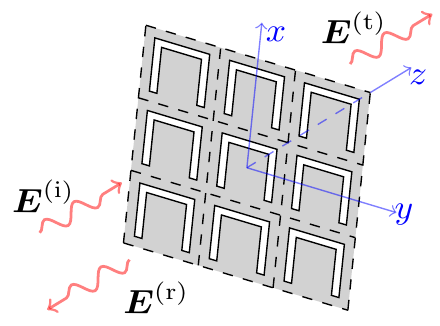}
\caption{
A periodic planar array with normally incident (i), reflected (r) and transmitted (t) waves.
}
\label{fig:screen}
\end{figure}

We consider the scattering of a linearly polarized electromagnetic plane wave by a periodically perforated metal screen in free space, see Figure~\ref{fig:screen}. The goal is to quantify the amount of transmitted power that passes through the structure and continues to propagate as a wave of the same frequency, polarization and direction as the incident wave. 
To accomplish this, we extend the initial theoretical results reported in~\cite{Gustafsson+etal2011}, and use them to impose a bandwidth bound on the power transmission of such structures. 
The theory is derived under the assumption that the structure is an infinitely thin two-dimensional periodic PEC screen of infinite extent in the plane normal to the incident wave direction.
In Sections~\ref{sec:Practice} and~\ref{sec:Measurements}, these assumptions are validated to be reasonable approximations for power transmission through real structures.

The screen is placed in the ${\rm xy}$-plane at $z=0$, and the unit cell is defined by the lattice vectors $l_{\rm x}\hat{\vec{x}}$ and $l_{\rm y}\hat{\vec{y}}$. 
The incoming wave with the associated electric field $\vec{E}^{\rm (i)}(k,\vec{r})=E^{\rm (i)}{\rm e}^{\iu \vec{k} \cdot \vec{r}}\hat{\vec{e}}$ is propagating in the positive $z$-direction, where $\hat{\vec{e}}$ is the polarization unit vector, {$k=\omega/\cu$ is the wave number in free space with $\omega$ and $\cu$ being the angular frequency and the speed of light in vacuum, respectively,} $\vec{k}=k\hat{\vec{z}}$ is the wavevector, $\vec{r}$ is the field position vector, and the time convention ${\rm e}^{-\iu\omega t}$ is used. Interaction between the incident wave and the structure gives rise to the scattered field. We denote the {scattered} field in $z<0$ as the reflected field $\vec{E}^{\rm (r)}(k,\vec{r})$, and the {total} field in $z>0$ as the transmitted field $\vec{E}^{\rm (t)}(k,\vec{r})$.
A spectral decomposition of the transmitted field in Floquet modes is
\begin{equation}
	\vec{E}^{({\rm t})}(k,\vec{r}) = \sum\limits_{m,n=-\infty}^{\infty} \vec{E}^{({\rm t})}_{mn}(k){\rm e}^{\iu \vec{k}_{mn} \cdot \vec{r}},
	\label{eq:floquet}
\end{equation}
where $\vec{k}_{mn}=k_{{\rm x},n}\hat{\vec{x}}+k_{{\rm y},m}\hat{\vec{y}}+k_{{\rm z,}mn}\hat{\vec{z}}$ 
are the modal wave vectors with ${k}_{{\rm x},n}=2\pi n/l_{\rm x}$, ${k}_{{\rm y},m}=2\pi m/l_{\rm y}$, $k_{{\rm z,}mn}=\sqrt{k^2-{k}_{{\rm x},n}^2-{k}_{{\rm y},m}^2}$ and $\vec{E}^{({\rm t})}_{mn}(k)$ are the expansion coefficients.
The latter are related to the incident field through a linear mapping 
\begin{equation}
	\vec{E}^{({\rm t})}_{mn}(k) = {\bf T}_{mn}(k)\cdot \vec{E}^{({\rm i})}(k,z=0),
	\label{eq:Et}
\end{equation}
where ${\bf T}_{mn}(k)$ are the transmission dyadic tensors. 
For frequencies below the first grating lobe, $f<c_0/\max \{l_{\rm x},l_{\rm y}\}$~\cite{Munk2000}, only the fundamental mode is propagating. We define the co-polarized transmission coefficient for the fundamental mode as $T(k) = \hat{\vec{e}}\cdot{\bf T}_{00}(k)\cdot\hat{\vec{e}}$. 

Given a transmission threshold $T_0$ we define the transmission bands as intervals of $k$, where $|T(k)|>T_0$. For the largest such interval (the main band) with endpoints $k_{\rm 1}$ and $k_{\rm 2}$, the fractional bandwidth is
\begin{equation}
B=2\frac{k_{\rm 2}-k_{\rm 1}}{k_{\rm 1}+k_{\rm 2}}.
\label{eq:B}
\end{equation}
In this paper, we characterize how the fractional bandwidth depends on various perforation shapes with respect to different metrics, such as aperture area, or size of a minimal enclosing square, see $S_{\rm p}$ and $a$ respectively, in Figure~\ref{fig:UC}. 
{\color{black} In the design examples, we strive towards having the bandwidth of the lowest-frequency transmission peak to be as close to the maximum attainable as possible.}

\section{Derivation of the sum rule}\label{sec:derivation}
In this section, the derivation of the sum rule is presented.
It is based on the passive properties of the screen~\cite{Zemanian1965}, 
with an associated system response that can be transformed into a
Herglotz function~\cite{Bernland+etal2011,IET2018}, 
associated with the scattering system.
An integral identity is applied to this function to obtain the extraordinary-transmission sum rule.
The main theoretical result is the sum rule in~\eqref{eq:sum_rule}, from which an upper bound of~\eqref{eq:B} is obtained in~\eqref{eq:sum_rule_B}.

\begin{figure}
\centering
   \begin{subfigure}[b]{0.45\linewidth}
   \includegraphics[width=\linewidth]{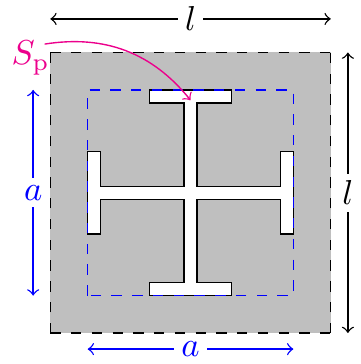}
   \caption{}
   \label{fig:UC}
   \end{subfigure}   
   \begin{subfigure}[b]{0.53\linewidth}
\includegraphics[width=\linewidth]{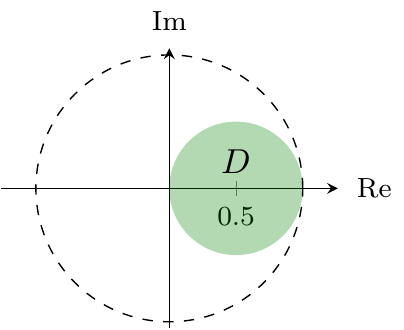}
\caption{}
\label{fig:T_complex}
\end{subfigure}
\caption{(a) An example of the unit cell geometry, with perforated area $S_\mrm{p}$, contained in a minimal enclosing square of size $a$, and the unit cell size $l=l_{\rm x}=l_{\rm y}$;
(b) Range of the transmission coefficient $T(k)\in D$ in the complex plane.
}
\end{figure}

Passivity of the scattering configuration~\cite{Zemanian1965,Nussenzveig1972,Gustafsson+etal2009} allows an analytical extension of $T(k)$ for $k\in\C^+$, where $\C^+=\{k\in\C: \Im k > 0 \}$ {is the upper half plane}.
Apart from analyticity, a few additional properties are required to construct a physical bound in the form of a sum rule. The impinging wave generates electric currents on the screen. From the assumption of negligible thickness of the screen it follows that the scattered field is symmetric relative to the screen, \ie $\vec{E}^{({\rm t})}-\vec{E}^{({\rm i})} =  \vec{E}^{({\rm r})}$ at $z=0$. 
This can be rewritten as $T(k)=1 + R(k)$, where $R(k)$ is the reflection coefficient defined for $\vec{E}^{({\rm r})}$ similarly as $T(k)$ is defined for $\vec{E}^{({\rm t})}$. This, combined with conservation of power $|T(k)|^2+|R(k)|^2\leq 1$, 
 yields
$\left|T(k)-1/2\right|\leq 1/2$. Thus, the transmission coefficient is a holomorphic mapping from the upper complex half-plane $\C^+$ to the closed disc $D$ with center at $1/2$ and radius $1/2$ in the complex plane, see the green disc in Figure~\ref{fig:T_complex}.

In order to obtain the sum rule, we transform the system response $T(k)$ in such a way that it becomes a Herglotz function, which is a mapping from the upper complex half-plane to 
{its closure}. Details about Herglotz functions and 
the associated integral identity used here are stated in Appendix~\ref{app:herglotz}.

Here, we consider a M\"{o}bius transform~\cite{Conway1973}, $m(\zeta)=\iu(1-\zeta)/\zeta$, which maps the 
disc $D$ to the closed upper complex half-plane. We compose it with the transmission coefficient to obtain a symmetric Herglotz function 
\begin{equation}
\label{eq:f_fun}
	g(k)=m ( T(k))=\iu\frac{1-T(k)}{T(k)}.
\end{equation}

A key element to derive the sum rule is the high and low frequency behavior of the transmission coefficient $T$, 
 see~\eqref{eq:Herglotzidentity} in Appendix~\ref{app:herglotz}. To determine the low-frequency behavior we utilize Babinet's principle: the field $\vec{E}^{({\rm t})}$ transmitted through an aperture screen and the field $\vec{E}_{\rm c}^{({\rm t})}$  transmitted through the complementary structure are related as $\vec{E}^{({\rm t})}+\vec{E}_{\rm c}^{({\rm t})}=\vec{E}^{({\rm i})}$~\cite{Landau+etal1984, vanBladel2007}, where $\vec{E}^{({\rm i})}$ is the incident field in both cases, see also a single aperture case in~\cite{Gustafsson2009}. Hence, the low-frequency expansion (\ie $k\toh 0$) can be found by investigating the complementary structure. In the complementary structure, the perforations are filled with perfect magnetic conductor (PMC) in the ${\rm xy}$-plane and the PEC is removed. The transmission coefficient of the complementary structure is $T_{\rm c}(k)=1-T(k)$~\cite{Gustafsson2009}. Its low-frequency expansion is~\cite{Kleinman+Senior1986,Gustafsson+Sjoberg2011,Gustafsson+etal2009,Sjoberg2009}
\begin{equation}
\label{eq:Tc}
T_{\rm c}(k)\sim 1 + \frac{\iu k \gamma}{2A} \qquad \text{as } k\toh 0,
\end{equation}
where $\gamma=(\hat{\vec{e}}\cdot \bm{\gamma}_{\rm e}\cdot\hat{\vec{e}} + (\hat{\vec{k}}\times\hat{\vec{e}})\cdot \bm{\gamma}_{\rm m}\cdot(\hat{\vec{k}}\times\hat{\vec{e}}))$, $\hat{\vec{k}}=\zvh$ is the wave propagation direction, $\bm{\gamma}_{\rm e}$ and $\bm{\gamma}_{\rm m}$ are the electric and the magnetic polarizability tensors of the complementary structure, respectively, and $A=l_{\rm x}l_{\rm y}$ 
is the area of the unit cell. This gives us the expansion for the perforated PEC screen
\begin{equation}
\label{eq:T}
	T(k)\sim -\frac{\iu k\gamma}{2A} \qquad \text{as } k\toh 0.
\end{equation}
Note that the polarizabilities used here are the polarizabilities for the complementary structure.
{Furthermore, for a planar PMC array and the electric field direction $\hat{\vec{e}}$ parallel to the array plane, the term $\hat{\vec{e}}\cdot \bm{\gamma}_{\rm e}\cdot\hat{\vec{e}}$ vanishes, and thus we only need to calculate the magnetic polarizability.
The magnetic polarizability of a PMC structure can be calculated as the electric polarizability of a PEC structure of the same shape, see \eg~\cite{Gustafsson+etal2009,Sjoberg+etal20101}.}

To construct the desired sum rule, we apply a Herglotz function with specific properties to $g(k)$ in~\eqref{eq:f_fun}. The resulting function is a Herglotz function, as non-zero Herglotz functions satisfy the property that a composition of two Herglotz functions is a Herglotz function~\cite{Bernland+etal2011}.
To obtain an effective sum rule,
we want to characterize the total attainable bandwidth. To do this, we need to {emphasize} the bands where the transmission is higher than a chosen threshold $T_0$ and to disregard the rest of the spectrum. The desired  
function $h_\varDelta$ should have the properties
$\Im h_\varDelta (g(k))=1$ when $|T|\geq T_0$ and zero otherwise.
The pulse Herglotz function~\cite{Bernland+etal2011}
\begin{equation}	
h_{\varDelta}(\zeta)=\frac{1}{\pi}{\ln}\frac{\zeta-\varDelta}{\zeta+\varDelta}
    \sim \begin{cases} \iu &\mbox{as } \zeta\to 0 \\
-\frac{2\varDelta}{\pi \zeta} & \mbox{as } \zeta\to\infty, 
\end{cases}
\end{equation}
satisfies these criteria and has previously been used to construct sum rules for passive metamaterials~\cite{Gustafsson+Sjoberg2010} and high-impedance surfaces~\cite{Gustafsson+Sjoberg2011}.
For any real-valued argument $x$ this function has the property $\Im h_{\varDelta}(x)=1$ for $|x|<\varDelta $ and $\Im h_{\varDelta}(x)=0$ for $|x|>\varDelta $.
{We use this property later to relate the resulting integral identity with the fractional bandwidth~\eqref{eq:B} for the lossless case.}
For the composed function $h_{\varDelta}(g(k))$, the connection between the parameter $\varDelta$ and the threshold $T_0$ is found from relating $\varDelta$ to a threshold value of $g(k)$ (i.e., when $|T(k)|=T_0$)
\begin{equation}
	\varDelta^2=\frac{1-T_0^2}{T_0^2}.
    \label{eq:Delta}
\end{equation}

Finally, we apply the {integral} identity~\eqref{eq:Herglotzidentity} to the function $h_{\varDelta}(g(k))$.
From~\eqref{eq:T} we obtain that 
$T(k)\sim -\iu k\gamma/(2A)$ for $k\toh 0$. 
Combining this result and the low-frequency asymptote of~\eqref{eq:f_fun}, we get $g(k)\sim -2A/(\gamma k)$ as $k\toh 0$.
Consequently, the function $h_{\varDelta}(g(k))$ has the low-frequency expansion 
$h_{\varDelta}(g(k))\sim  k \gamma \varDelta /(A\pi)$ for $k\toh 0$.
Performing the same steps for the high frequency limit yields $h_{\varDelta}(g(k))\sim o(k)$ as $k \toh\infty$.
Thus, according to~\eqref{eq:Herglotzidentity} we find the sum rule
\begin{equation}
\label{eq:sum_rule_k}
	\int \limits_{0}^{\infty}\dfrac{\Im h_{\varDelta}(g(k))}{k^2}\diff k={\frac{\gamma\varDelta}{2A}}.
\end{equation}
After substituting $\lambda = 2\pi/k$ and reusing $g(\lambda)$ for~\eqref{eq:f_fun} as a function of wavelength,
 an alternative form of the sum rule is
\begin{equation}
\label{eq:sum_rule}
	\int \limits_{0}^{\infty}\Im h_{\varDelta}(g(\lambda)) \diff\lambda
    ={\frac{\gamma\varDelta\pi}{A}}.
\end{equation}
From this sum rule expression we deduce the upper bound of~\eqref{eq:B} convenient for practical use.
The sum rule shows that the total sum of transmission bands of an aperture array is determined by the array's polarizability per unit area.
Note that the right-hand side of~\eqref{eq:sum_rule} is always strictly positive
and hence there must exist intervals with non-zero transmission. Moreover, 
the transmission is perfect{\color{black}{,}} $|T(\lambda_0)|=1${\color{black}{,}} for some wavelength $\lambda_0$ if the structure is resonant below the onset of grating lobes and the cross polarization is negligible. This is a consequence of a lossless scattering system {\color{black}{with}} $|T|^2+|R|^2 = 1$, for which $T$ is located on the boundary circle of $D$, in Figure~\ref{fig:T_complex}. This implies  that $\Im h_{\varDelta}(g(\lambda))=1$ for some wavelength interval of nonzero length, \ie there always exists a transmission band with an arbitrarily high level of transmission.

{Note that, although ohmic losses are eliminated for screens made of PEC material, the scattering system is in general lossy due to radiation in other modes than the co-polarized fundamental mode in~\eqref{eq:Et}.  
Such radiation is perceived as losses from the system point of view, and includes higher-order modes radiating above the grating lobe frequency, as well as the cross-polarized mode below the first grating lobe. }  
We refrain here from considering lossy materials {\color{black} and impedance surfaces} from a theoretical perspective,
as the resulting lossy case bound is in general not tight. Instead, we treat our lossless PEC model as an approximation of a highly conductive low-loss screen. Further discussion on the validity of the model is provided in Sections~\ref{sec:Practice}-\ref{sec:Measurements}.

\begin{figure}
\begin{center}
\includegraphics[width=\linewidth]{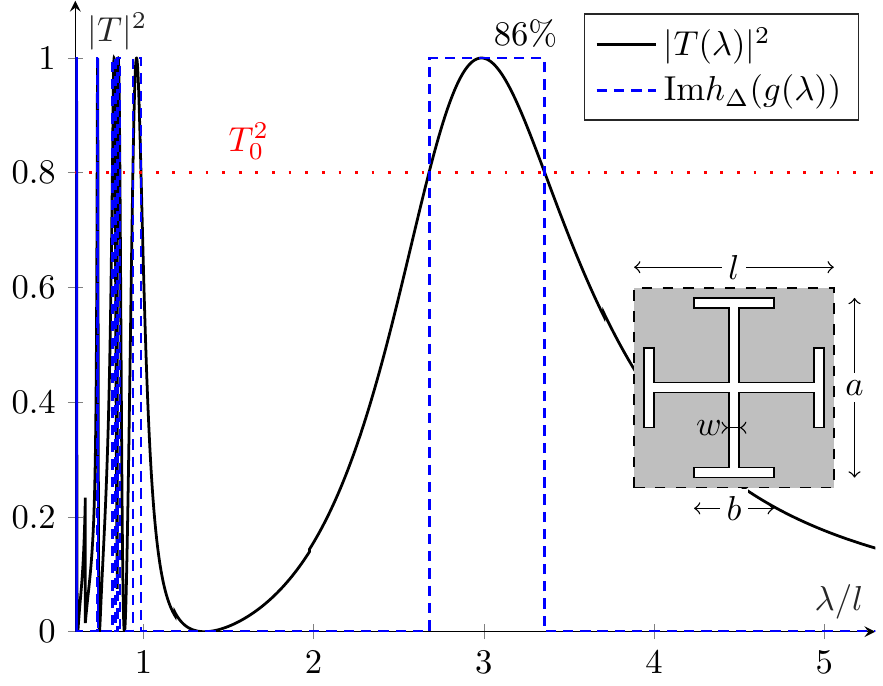}
\end{center}
\caption{Cross potent: transmittance as a function of wavelength and the unit cell geometry. Transmission bands with respect to the {\color{black} transmittance} threshold level $T_0^2=0.8$ (dotted line) are shown by the integrand function $\Im h_{\varDelta}(g(\lambda))$ of ~\eqref{eq:sum_rule}.
}
\label{fig:JC}
\end{figure}

For practical applications the integration over a finite interval of wavelengths $[\lambda_{\rm a},\lambda_{\rm b}]$ is performed (\eg see Figure~\ref{fig:JC} with $[\lambda_{\rm a},\lambda_{\rm b}]=[0.6l,5.2l]$, where $l$ is the unit cell size)
\begin{equation}
\label{eq:sum_rule_finite}
	\int \limits_{\lambda_{\rm a}}^{\lambda_{\rm b}}\Im h_{\varDelta}(g(\lambda))\diff\lambda
    \leq {\frac{\gamma\pi\varDelta}{A}}.
\end{equation}
Assume now that within the interval $[\lambda_{\rm a},\lambda_{\rm b}]$ there are a number of mutually disjoint subintervals, 
where $|T|\geq T_0$. 
{As an example, in Figure~\ref{fig:JC} we observe two intervals for $\lambda/l>0.9$ with transmittance higher than $T_0^2=0.8$, 
where the widest is located around $\lambda/l=3$.}
In this paper, we focus mainly on the bandwidth of the widest transmission band {\color{black} 
even though the sum rule includes all the transmission windows}.
If we retain only the contribution of the largest transmission band with endpoints $\lambda_1$ and $\lambda_2$, and normalize~\eqref{eq:sum_rule_finite} with the central wavelength $\lambda_0=(\lambda_1+\lambda_2)/2$  of the corresponding band, we obtain a bound for the fractional bandwidth
\begin{equation}
\label{eq:sum_rule_B}
	B = 2\dfrac{\lambda_{\rm 1}-\lambda_{\rm 2}}{\lambda_{\rm 1}+\lambda_{\rm 2}}\leq \frac{\gamma\pi\varDelta}{A\lambda_0}.
\end{equation}
{\color{black} 
Note that due to the relation $|T|^2+|R|^2 = 1$ below the first grating lobe frequency and assuming negligible cross polarization, the numerical results can be equivalently presented in the reflectance.
We, however, focus on transmission bandwidth in this paper.}

\section{Numerical examples and applications of the bound}
\label{sec:num}
We begin this section with illustrating the sum rule~\eqref{eq:sum_rule} by analyzing a given FSS design.
The numerical example in Figure~\ref{fig:JC} shows the transmittance $|T|^2$ through an array of {cross-potent} (sometimes referred to as Jerusalem cross)~\cite{Mittra+etal1988,Munk2000} shaped apertures as a function of the normalized wavelength. The assumptions of the idealized model are retained: the screen is infinitely thin and made of PEC material, and the array of perforations is infinitely periodic. The unit cell geometry is given, with $l_{\rm x}=l_{\rm y}=l$, slot width $w=l/20$, and parameters $a=0.9l$ and $b=0.4l$. The numerical analysis was performed in CST MW Studio using the frequency domain solver and Floquet mode ports.
The results show the main transmission band (transmittance threshold level $T_0^2=0.8$) centered at $\lambda=2.9l$, with the fractional bandwidth $B=0.24$. This accounts for $86\% $ of the upper bound limit in~\eqref{eq:sum_rule_B}. For the wavelengths shorter than $0.9\lambda$, we observe multiple narrower peaks. Due to the resonance nature of the phenomenon, an infinite number of such peaks is expected in the short wavelength limit. The contributions from these peaks, along with the grating lobes, are also accounted for in the left hand side of the sum rule~\eqref{eq:sum_rule}.

In the above cross-potent example, the sum rule is used to analyze a given FSS design. Additionally, the sum rule is instrumental to design and optimize FSS, which we discuss and illustrate in the remainder of this section.
One of the most crucial performance parameters of a periodically perforated screen is the frequency bandwidth over which the screen is transparent. 
The bandwidth optimization of periodic screens typically involves a considerable amount of full-wave numerical simulations in order to tune the design. Thus, tools that guide the optimization process and reduce the number of simulations are desired. The sum rule~\eqref{eq:sum_rule_B} is such a tool, as it provides a quantitative estimate of the total attainable bandwidth. This can be used in two ways. First, the total attainable bandwidth is bounded by the static polarizability of a perforation element according to~\eqref{eq:sum_rule_B}.
Testing the static polarizability for each design candidate can thus replace numerically costly wide-frequency-range full-wave simulations in the search of preliminary structure.
Second, the total attainable bandwidth, obtained from the polarizability, can serve as a reference for the fraction of the total bandwidth in the main frequency band. Using this reference, we can optimize the main frequency band to utilize most of the physically attainable bandwidth. In this section, we demonstrate an optimization procedure by maximizing the frequency bandwidth of a transmission window through a perforated screen, while keeping the area of perforations $S_{\rm p}$ low ($1-5\%$ of the total screen area). In general, different metrics can be considered instead of $S_{\rm p}$, for example, the size of the smallest enclosing square.

\begin{figure}
\begin{center}
\includegraphics[width=\linewidth]{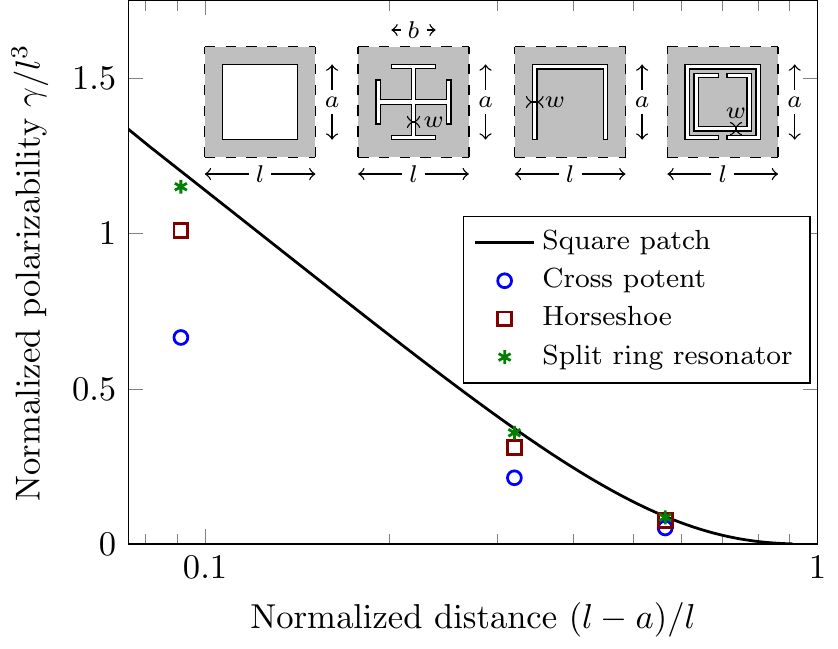}
\end{center}
\caption{Normalized polarizability $\gamma/l^3$ of infinitely thin PEC periodic structures as a function of the normalized distance between two adjacent square patches, $l\in \{1.1a,~1.47a,~2.29a\}$. 
{The external applied field for polarizability calculation is directed vertically with respect to the unit cells in the inset.}
}
\label{fig:gamma_dist}
\end{figure}

The total achievable bandwidth is determined by the polarizability of the corresponding complementary structure according to~\eqref{eq:sum_rule_B}, as discussed above. We use this as a guideline to choose the preliminary perforation design.
Figure~\ref{fig:gamma_dist} compares the normalized polarizability $\gamma/l^3$
of an array of square PEC patches of size $a\times a$ and period $l_{\rm x}=l_{\rm y}=l$ with periodic PEC arrays, tightly enclosed by the square patch structure. Three shapes of enclosed unit cell designs are considered: a cross potent, a horseshoe and a split ring resonator. According to the monotonic growth of polarizabilities with volume, the polarizability of an enclosed object cannot exceed the polarizability of an enclosing object~\cite{Schiffer+Szego1948,Sjoberg2009}. Thus, the polarizability of the square patches is the upper bound for the enclosed designs. We observe in Figure~\ref{fig:gamma_dist} that the horseshoe and the split ring designs approach the upper bound, and thus make a good use of the unit cell geometry. The split-ring-resonator unit cell outperforms the horseshoe-shaped design when they are compared with respect to normalized distance $(l-a)/l$ between the adjacent perforations.

\begin{figure}
\begin{center}
\includegraphics[width=\linewidth]{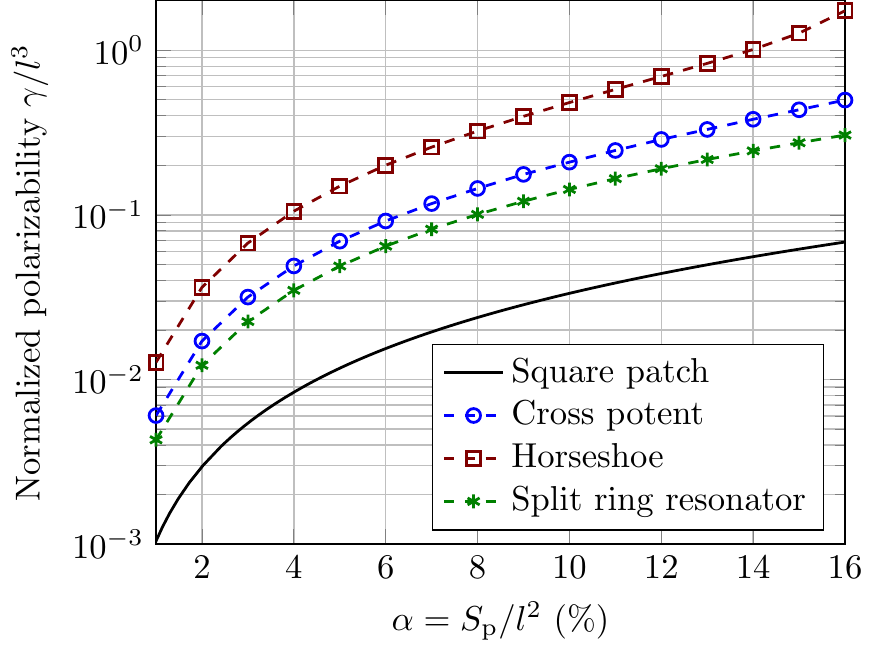}
\end{center}
\caption{Normalized polarizability $\gamma/l^3$ of infinitely thin PEC periodic structures as a function of perforated area.
{The external applied field for polarizability calculation is directed vertically with respect to the unit cells in the inset of Figure~\ref{fig:gamma_dist}.}
}
\label{fig:gamma}
\end{figure}

\begin{figure}
\begin{center}
\includegraphics[width=\linewidth]{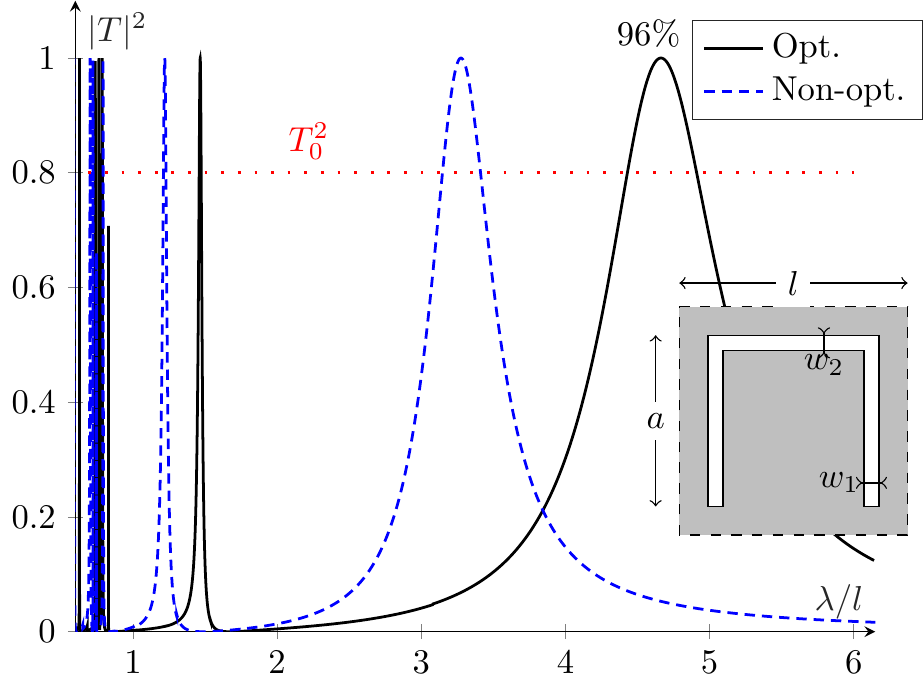}
\end{center}
\caption{Horseshoe transmittance, optimized(black curve) and non-optimized(dash-dotted) designs. In both cases $\alpha = S_\mrm{p}/A=5\%$.}
\label{fig:horseshoe}
\end{figure}

As an alternative evaluation of performance, we can investigate how the shapes perform with respect to the perforation area.
Figure~\ref{fig:gamma} shows the normalized static polarizability $\gamma/l^3$ as a function of the percentage of perforation area in the total area of the screen $\alpha = S_\mrm{p}/A$, where $l=l_{\rm x}=l_{\rm y}$ 
is the size of a unit cell, see Figure~\ref{fig:UC}.
Here, solid, $\circ$-dashed, $\square$-dashed and $\star$-dashed lines correspond to square hole, {cross potent}, horseshoe and split ring resonator designs, respectively. All the designs were contained within the square of size $a$, and the slot width for the {cross potent}, horseshoe and split ring was fixed at $w=a/17$, see \eg Figures~\ref{fig:JC} and \ref{fig:horseshoe}, while the unit cell size $l$ was varied in the range $[ 1.03a,11.15a ]$, $[ 1.04a,4.67a ]$, $[ 1.03a,4.12a ]$ and $[ 1.38a,6.17a ]$ 
for the corresponding design, respectively. The evaluation of static polarizabilities was performed via a variational approach~\cite{Sjoberg2009} in COMSOL Multiphysics electrostatic solver. 

We observe among all considered shapes that the horseshoe utilizes the perforation area better than the other shapes, in the sense of the upper bound $\gamma/l^3$ of the total attainable bandwidth. 
The square hole perforations are given as a reference, and all the suggested designs outperform it. The same total bandwidth, as achieved by cutting out $15\%$ of the screen with square-shaped perforations, can be attained by cutting out only $2.5\%$ of the screen with the horseshoe-shaped perforations.
Additionally, the horseshoe-shaped perforations have better mechanical stability compared to the other considered designs, which finalizes the choice of the preliminary structure.

Having chosen a horseshoe design as a preliminary structure, we perform optimization of its geometrical parameters with respect to its transmission bandwidth.
Consider the following optimization problem. For a given upper limit $\alpha_0$ of $\alpha=S_\mrm{p}/A$, make the fractional bandwidth $B$ as close as possible to its upper bound given by the right hand side of~\eqref{eq:sum_rule_B}. We denote the ratio between the bandwidth and its upper bound as $\eta(\Omega,\varDelta)=B/(\gamma \pi \varDelta /A\lambda_0)$.  
The optimization problem is formulated as follows for a given~$\varDelta$
\begin{equation}
\begin{aligned}
& \mathop{\text{maximize}}_{\Omega}
& & \eta(\Omega,\varDelta) \\
& \text{subject to}
& & \alpha(\Omega) \leq \alpha_0,
\end{aligned}
\label{eq:optimization}
\end{equation}
where the optimization is performed over the parametrized geometry $\Omega$ of the aperture. We use a genetic algorithm optimization for geometric parameters of a horseshoe shaped aperture (the shape choice is motivated by Figure~\ref{fig:gamma}). 

Figure~\ref{fig:horseshoe} shows the results of optimization of the horseshoe perforation geometry with $\alpha_0=5\%$. We start with a non-optimized design given by the size $a$, and
$l=1.69a$, $w_1=w_2=0.049a$. Optimization~\eqref{eq:optimization} yields the design given by $l=1.43a$, $w_1=0.049a$, and $w_2=0.0047a$. We observe that the bandwidth is improved approximately twice, and the main peak contains $96\% $ of the attainable bandwidth, according to~\eqref{eq:sum_rule_B}.

\section{Implementation of real structures}
\label{sec:Practice}

\begin{figure}[!h]
\includegraphics[width=\linewidth]{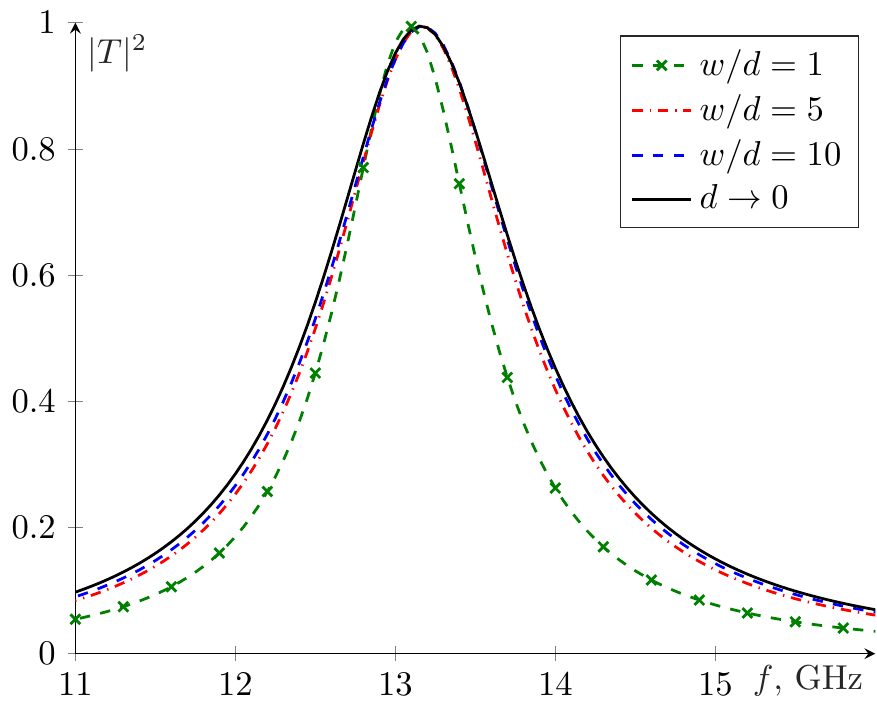}
\caption{{Simulated transmittance} through an array of horseshoe apertures for different ratios of the slot width to the screen thickness. } 
\label{fig:thin_vs_thick_freq}
\end{figure}

In Sections~\ref{sec:Theory1}-\ref{sec:num}, the theoretical and numerical evaluation of the screen were performed under certain assumptions. Here, we investigate the validity of these results when the assumptions are relaxed. One of the key assumptions was that the screen is infinitely thin. Figure~\ref{fig:thin_vs_thick_freq} illustrates how the screen thickness $d$ affects the screen's transmission characteristics. We consider transmission through a screen with horseshoe-shaped apertures, with slot width $w_1=w_2=w$, see Figure~\ref{fig:horseshoe} for the unit cell geometry. The transmission through an infinitely thin screen is compared with three screens with width-to-thickness ratios $w/d=\{1,5,10\}$. For $w/d=1$, we observe a noticeable bandwidth reduction in comparison with the infinitely thin case. However, when $w/d=10$, the difference between the transmittance of the infinitely thin screen and the screen of thickness $d$ is negligible, resulting in a bandwidth reduction of about $2\%$ (with the threshold $T_0^2=0.8$). 
Figure~\ref{fig:thin_vs_thick_freq} shows that the transmission bandwidth is reduced with decreasing $w/d$ ratio. This implies that the inequality in~\eqref{eq:sum_rule_B} is still valid for cases with a finite thickness. However, when the slot width becomes comparable to the slot thickness, the bound is not tight.

\begin{figure}[!h]
\includegraphics[width=\linewidth]{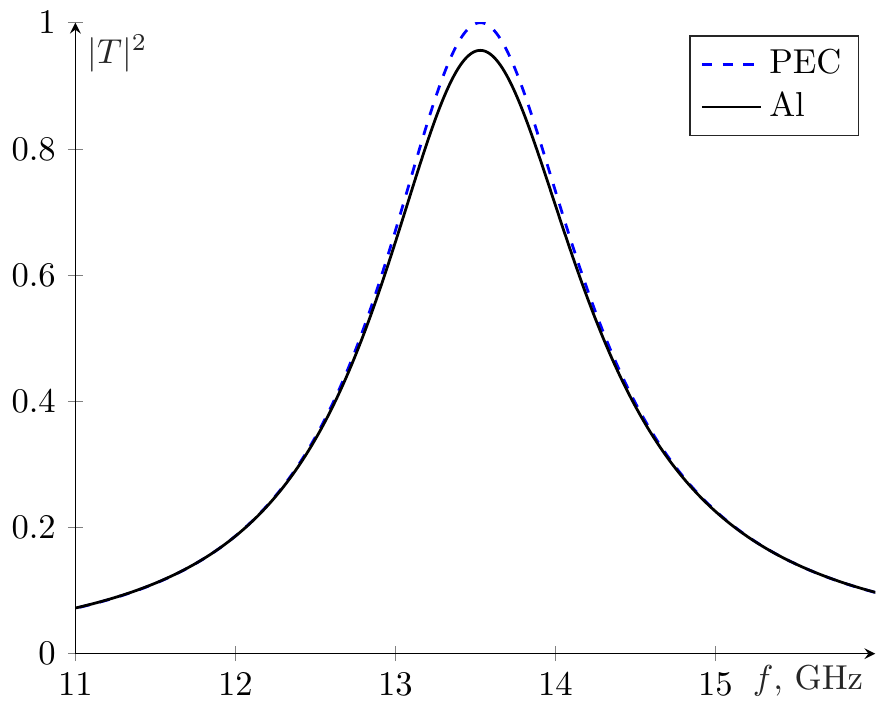}
\caption{Simulated transmittance of the horseshoe design for manufactured sample, PEC and Al comparison.
}
\label{fig:pecalreal}
\end{figure}

The second crucial assumption made in the derivation of the sum rule was the PEC material of the screen. Therefore, candidates for screen material should be highly conductive low-loss metals. To reconcile this requirement with limitations put on thickness and mechanical stability,  aluminum foil was chosen. 
Alternative options were metalized dielectric substrate, copper sheet and silver foil. However, these options impose  issues which are hard to resolve in the sum rule or fabrication. 
Figure~\ref{fig:pecalreal} shows the simulated transmittance for a perforated screen made of PEC or aluminum. {The geometrical parameters of the screen are the same as of the manufactured sample, {to be} discussed in the next section.} The aluminum screen has slightly lower amplitude (about 5\%) in the transmission peak in comparison with the PEC screen. However, the bandwidth reduction is negligible. Thus, the sum rule is applicable to aluminum screens and it is relatively tight. 

In the sum rule we also considered an infinite periodicity of the screen. Ref.~\cite{Camacho+etal2016} reports that 30 periods in both dimensions of the screen is sufficient to ensure a negligible difference in transmission between finite and infinite structures. The edge effects can be compensated by time-gating.

\begin{figure}
\centering
   \includegraphics[width=\linewidth]{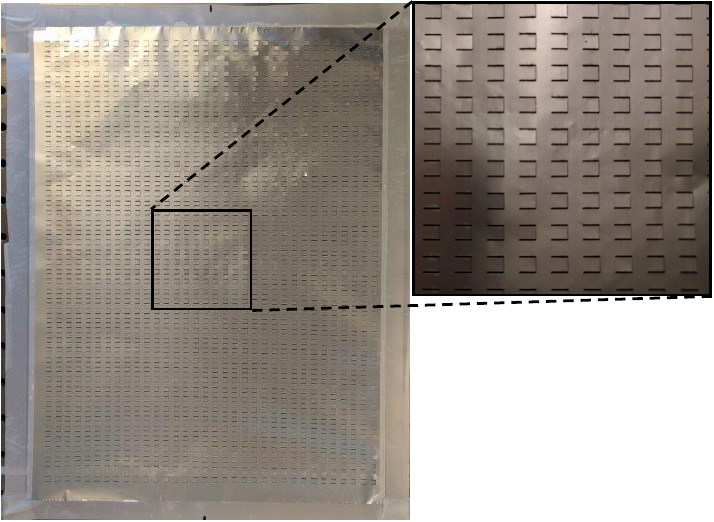}
\caption{Sample manufactured in aluminium foil: the entire sample of the size 238\unit{mm}$\times$320\unit{mm} and a close up of the manufactured horseshoe design. 
}
\label{fig:samples}
\end{figure}

\section{Measurements}\label{sec:Measurements}

\begin{figure}
\begin{center}
\includegraphics[width=\linewidth,trim={0 0 0 6cm},clip]{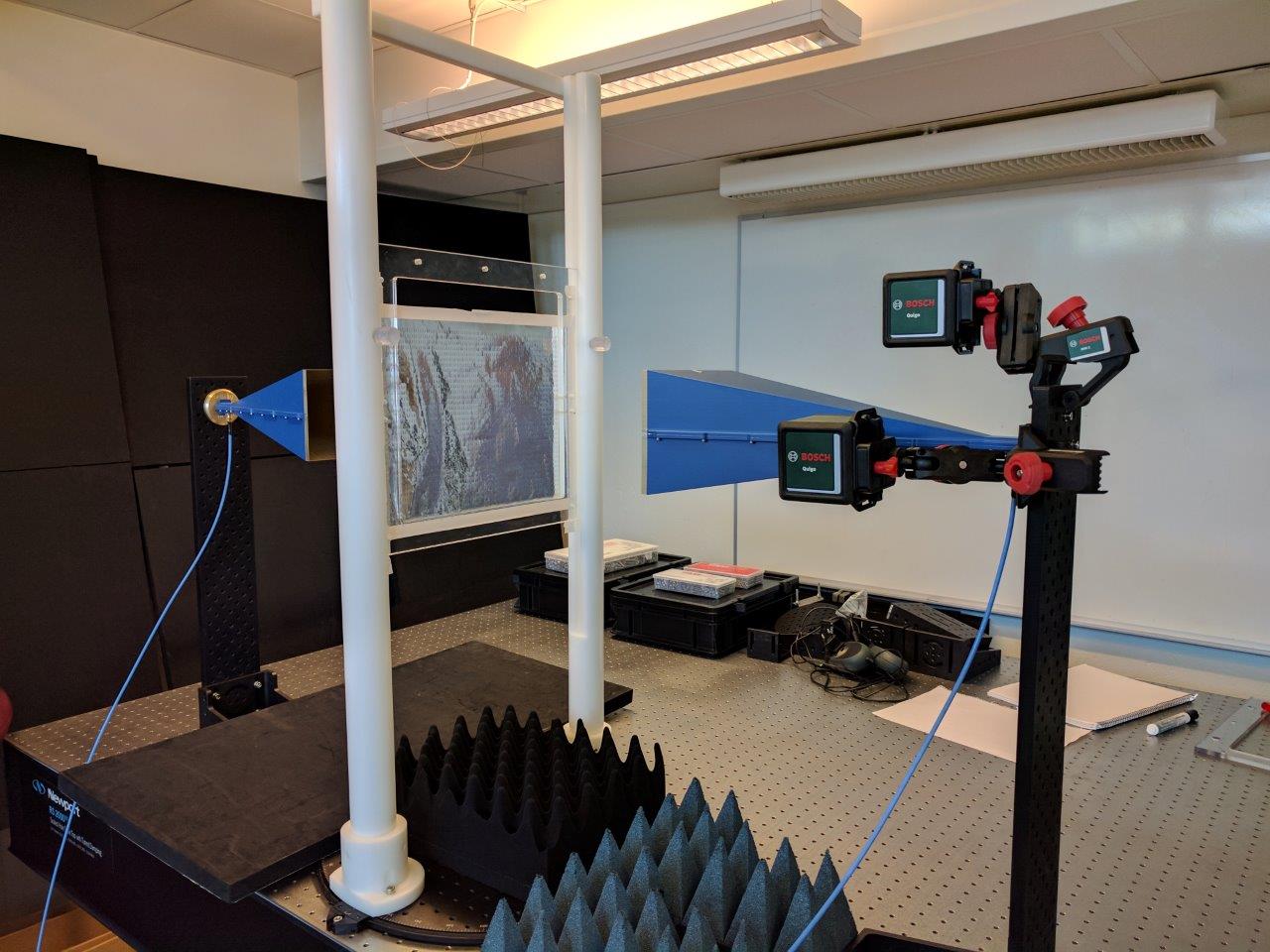}
\end{center}
\caption{Measurement setup mounted on an optical table. Two blue standard gain, horn MVG SGH1240 antennas were used as both receiver and transmitter. The horns were aligned using two Bosch Quigo Cross line lasers, seen here mounted on the right antenna. The sample was mounted on a custom made polymethyl methacrylate frame and held up by two plastic stands.
}
\label{fig:measurements}
\end{figure}

The final manufactured sample had the unit cell geometry given by the inset in Figure~\ref{fig:horseshoe} with $l=6.57$\unit{mm}, $a=3.43$\unit{mm}, $w_1=0.3$\unit{mm}, and $w_2=0.06$\unit{mm}. The aperture array was laser milled by a ProtoLaser U3 machine in a sheet of aluminum foil of thickness $d=0.018$\unit{mm}.
The array consisted of $34\times 45=1530$ apertures, and $\alpha=5\%$; see Figure~\ref{fig:samples} for the manufactured sample.

The measurement setup is shown in Figure~\ref{fig:measurements}. The sample was fixed in a polymethyl methacrylate (PMMA) 
frame fastened by two plastic stands equidistant to the transmitting and receiving antennas. Standard gain horn Satimo SGH1240 antennas were used, with the nominal frequency range $12.4-18.0$\unit{GHz}. The antennas were installed at the distance of 1\unit{m} from each other. The reference transmission measurements were performed with the empty PMMA frame instead of the sample and multipath reflections from the surrounding objects and surfaces were filtered out in the time domain by using time-gating~\cite{Ericsson+etal2017,Bryant1993} 
utilizing a tapered cosine window. 

\begin{figure}
\begin{center}
\includegraphics[width=\linewidth]{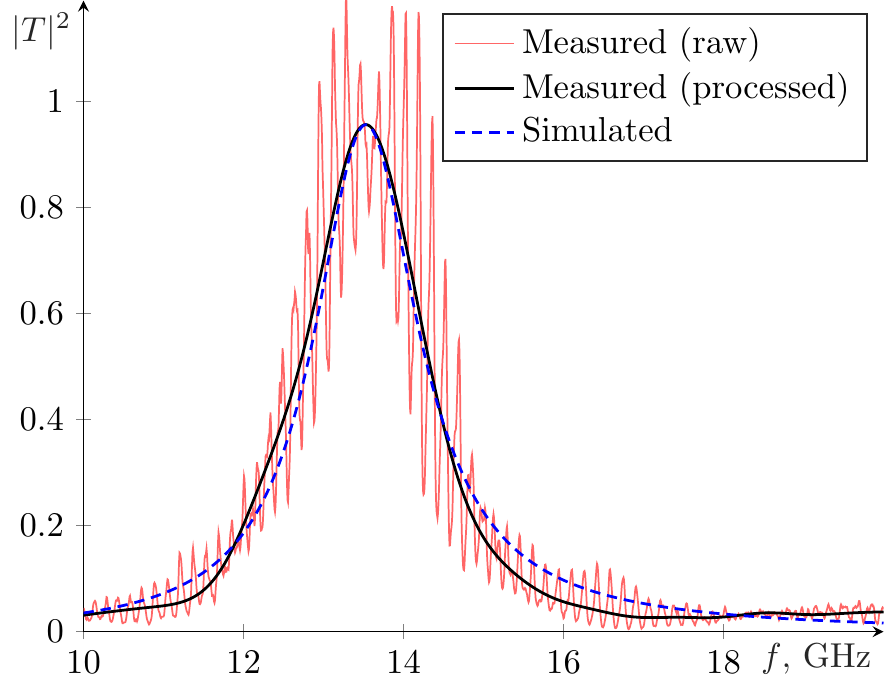}
\end{center}
\caption{Horseshoe slot array: comparison between measured raw data (red), processed data (black solid) and simulated (blue dashed) transmittance. }
\label{fig:results}
\end{figure}

The transmission through the manufactured sample was measured in the frequency range $10-20\,\unit{GHz}$ to capture the first transmission peak and to filter multipath components in an efficient manner.
In Figure~\ref{fig:results}, the red and black solid curves correspond to the raw data of the measured transmittance 
and the processed data obtained by filtering out multipath 
propagation components 
using time gating, respectively, and the dashed curve corresponds to the simulated transmittance of the infinitely-periodic model of the sample.
We observe a fine agreement between the measured and simulated transmittances in the whole frequency range.
The magnitude and the frequency of the resonance perfectly coincide for simulation and measurements.

The optimized PEC-bandwidth of the lowest-frequency peak, as shown in Section~\ref{sec:num}, reaches 96\% of the available physical bandwidth, based on the sum rule utilizing the polarizability of the perforation 
~\eqref{eq:sum_rule_B}. As we saw in Section~\ref{sec:Practice}, a finite thickness, but small in comparison with the perforation size, together with a finite but high conductivity made small perturbations to the transmission peak. 

By comparing the measured result with the PEC-simulated results at the 80\% transmittance threshold level, we find that the time-gated measured transmission peak has 98\% of the available bandwidth of a PEC-based structure.
The measured transmission peak is centered at the frequency of 13.52~\unit{GHz} with the fractional bandwidth of 5.83\%.

The remarkable agreement between measured and PEC-simulated results validate the use of the PEC-based sum rule as a tool to predict the physically maximum available bandwidth in thin and highly conductive EoT-screens. We further note that the PEC-based upper bound solely utilizes the observation that the screen is a passive system.

\section{Conclusions}\label{sec:Conclusions}
In this paper we have derived a sum rule for periodic structures and applied it to an FSS design problem. We have shown numerically, that a periodic PEC-like infinitely thin screen with  5\% of its total area cut out as horseshoe shaped perforations can have up to 96\% 
of its physically attainable bandwidth in its largest transmission window. The transmittance threshold of this study was set to 80\%. 
Our numerical investigations illustrate that small perturbations of the PEC-screen accounting for a finite thickness and a finite but high conductivity marginally perturbed the transmission result. This indicates the validity of the sum rule for real applications, even though it was derived for an ideal model. 

We have experimentally validated our results by showing that the transmission characteristics of the first transmission window of a horseshoe design, optimized with the use of the sum rule, fabricated in a 0.018$\unit{m m}$ thick highly conducting aluminum foil with horseshoe perforations, accurately matches the corresponding simulations.
The mutual agreement between the theoretical limitations, numerical and experimental validation is high. 
The choice of frequency band was selected to fully utilize the range of the experimental equipment. We conclude that the sum rule can be used to predict the results of 
transmission experiments with highly conductive metal films in the $\unit{GHz}$ range, and may be of use in understanding the phenomena at other frequencies. We have also observed that the perforation shape needed to maximize the performance of this phenomenon can be rather simple and still gather a high degree of transmission in one transmission window.   

The theoretical sum rule result~\eqref{eq:sum_rule} shows that a transmission band at long (in comparison to the periodicity of the structure) wavelengths exists for any type of perforations, even infinitely small ones ($\alpha\to 0$). However, the bandwidth of the transmission peak is proportional to the polarizability, closely related to the shape and size of the perforations. 
As a result, it is shown that the static polarizability, and hence, the transmission bandwidth of an array of square apertures can be attained by periodic perforations of much smaller relative area $\alpha$, see  Figures~\ref{fig:gamma_dist}~and~\ref{fig:gamma}.
The sum rule~\eqref{eq:sum_rule} implicates that in the ideal setting the ratio $T/\alpha$ can be infinitely large.

The good agreement between the measured transmission peak and the corresponding PEC simulations was enabled by a careful choice of material. By utilizing aluminum foil we stayed relatively close to the idealized PEC case. There was no dielectric material supporting the metal, and the foil had high conductivity, which ensured a high value of transmittance in the transmission peak. The foil was also thinner than the smallest slot in the design. This meant that there was no waveguide-like phenomenon occurring in the slots. Such an effect has a tendency to shift the spectral localization of resonances. This can be compared to the initial investigation performed at optical frequencies~\cite{Ebbesen+etal1998,Koerkamp+etal2004,Moreno+etal2001}, where the aperture sizes are small compared to the thickness of the materials they are etched in. This explains why these studies exhibit deviations from ideal models.

The derived sum rule~\eqref{eq:sum_rule} provides an upper bound on the bandwidth, which determines the largest attainable bandwidth for any aperture, enclosed in \eg a square of a given size. Additionally, it allows us to  verify that the total bandwidth for a given aperture cannot be improved by  redesigning the aperture within the enclosing square.
In this paper, we have shown an example of how the sum rule can be used in the design process of a perforated screen. 
Given maximal transmission bandwidth as the design goal, while having constrained fractional aperture area, the sum rule serves as an upper limit to optimize towards. However, other constraints and optimization goals are possible.
The optimization problem might be formulated as maximization of the bandwidth of transmission peaks at wavelengths closest to the array period. Here, the sum rule can serve as a reference to maximally attainable bandwidth.

An interesting continuation of this work might be a theoretical investigation of transmission through infinitely-thin perforated impedance surface.
However, the approach of this paper does not directly translate to the impedance-surface case. The impedance surface does not have zero transmission in low-frequency limit, which is one of the key elements used in the derivation of the presented sum rule, and therefore requires further investigation.
One of the possible alternatives here is to use numerical techniques, approximating the system function with a linear combination of Herglotz functions \cite{Ivanenko+etal2019a,Nordebo+etal2014b}.

\appendices

\section{Herglotz functions}
\label{app:herglotz}
\setcounter{equation}{0}
\numberwithin{equation}{section}
A Herglotz function is a holomorphic function $h(\zeta)$ such that $\Im h(\zeta)\geq 0$ whenever $\Im \zeta>0$, \ie it is a mapping from the upper complex half plane to its closure.
Functions of this class can have a family of integral identities~\cite{Bernland+etal2011}, also known as sum rules. 

Consider a Herglotz-function such that
\begin{equation}
	h(\zeta) = \begin{cases} a_{-1}\zeta^{-1}+a_1\zeta+o(\zeta) &\mbox{as } \zeta\hat{\rightarrow}0, \\
    b_1\zeta +  o(\zeta^{-1}) &\mbox{as } \zeta\hat{\rightarrow}\infty,
    \end{cases}
\end{equation}
where the coefficients $a_{-1}$, $a_1$ and $b_1$ are real-valued. Here $\zeta=x+\iu y$. A sum rule~\cite{Bernland+etal2011,IET2018} for the Herglotz function $h$ with the above expansion is:
\begin{multline}\label{eq:Herglotzidentity}
  \frac{2}{\pi} \int \limits_{0+}^{\infty}  \frac{\Im h(x)}{x^{2}} \diff x
  \stackrel{\rm def}{=} 
  \lim_{\varepsilon\rightarrow 0+}\lim_{y\rightarrow 0+}\frac{2}{\pi}\int \limits_{\varepsilon}^{1/\varepsilon}\frac{\Im h(x+\iu y )}{x^{2}}\diff x \\
  =a_{1}-b_{1}.
\end{multline}

Above, $\hat{\rightarrow}$ denotes the limit in a cone  $\alpha\leq\arg(\zeta)\leq(\pi-\alpha)$ for some $\alpha>0$. 
Throughout this paper, we utilize the symmetry $h(\zeta)=-h^*(-\zeta^*)$, which follows from the real-valuedness of the function in the time domain.

\section*{Acknowledgment}
We are grateful to Martin Nilsson for aiding in the sample manufacturing process.
The authors acknowledge the support of the Swedish foundation for strategic research (SSF) under the grant `Complex analysis and convex optimization for electromagnetic design'. 



\newpage

\begin{IEEEbiography}[{\includegraphics[width=1in,height=1.25in,clip,keepaspectratio]{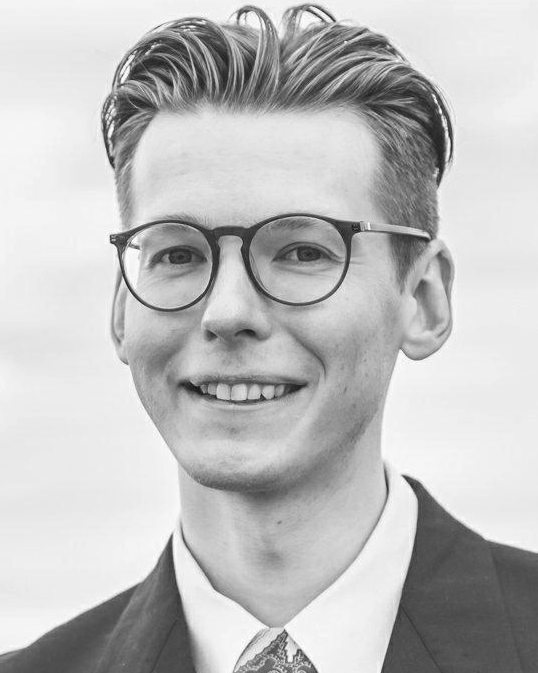}}]{Andrei Ludvig-Osipov} received his B.Sc. and M.Sc. degree in Radio Science and Telecommunications from St.~Petersburg State Electrotechnical University, Russia, in 2012 and 2014 respectively. 
He was a research engineer at R\&D Institute of Radio Science and Telecommunications, St.~Petersburg, Russia in 2013-2015.
Since 2015 he pursues a Ph.D. degree at the Department of Electromagnetic Engineering, KTH Royal Institute of Technology, Stockholm, Sweden.
His research interests include fundamental bounds in electromagnetics, stored energies, scattering and antenna theory applied to periodic structures, as well as signal processing.
\end{IEEEbiography}

\begin{IEEEbiography}[{\includegraphics[width=1in,height=1.25in,clip,keepaspectratio]{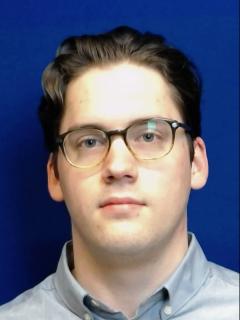}}]{Johan Lundgren}
received his M.Sc. degree in engineering physics from Lund University, Sweden, in 2016. He is currently working towards his Ph.D. degree with the Electromagnetic Theory Group at the Department of Electrical and Information Technology, Lund University. His research interests are in electromagnetic scattering, periodic structures, electromagnetic properties of materials and wave propagation.
\end{IEEEbiography}

\begin{IEEEbiography}[{\includegraphics[width=1in,height=1.25in,clip,keepaspectratio]{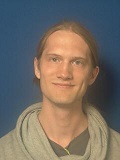}}]{Casimir Ehrenborg (S'15)} 
received his M.Sc. degree in engineering physics  from Lund University, Sweden, in 2014. He is currently a Ph.D. student in the Electromagnetic Theory Group, Department of Electrical and Information Technology, Lund University. In 2015, he  participated in and won the IEEE Antennas and  Propagation Society Student Design Contest for his body area network antenna design. In 2019, he was awarded the IEEE AP-S Uslenghi Letters Prize for best paper published in IEEE Antennas and Propagation Letters during 2018. His research interests include small antennas, stored energy, phase and radiation centers, as well as physical bounds.
\end{IEEEbiography}

\begin{IEEEbiography}[{\includegraphics[width=1in,height=1.25in,clip,keepaspectratio]{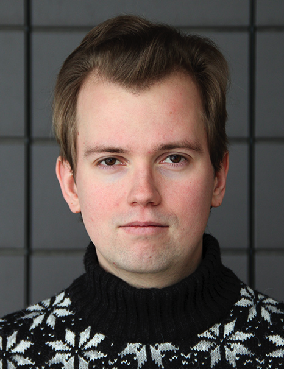}}]{Yevhen Ivanenko} (S'16) received his B.Sc. degree in radio electronic devices and M.Sc. degree in electronic instruments and devices from Odessa National Polytechnic University, Ukraine, in 2011 and 2013, respectively, as well as M.Sc. degree in electrical engineering and Lic. degree in physics from Linn\ae us University, Sweden, in 2015 and 2018, respectively. Currently, he is working towards his Ph.D. degree in Waves, Signals and Systems group at the Department of Physics and Electrical Engineering, Linn\ae us University. His research interests include electromagnetic properties of materials, composite materials, as well as optimal performance bounds and realizations for passive and non-passive electromagnetic systems. 
\end{IEEEbiography}

\begin{IEEEbiography}[{\includegraphics[width=1in,height=1.25in,clip,keepaspectratio]{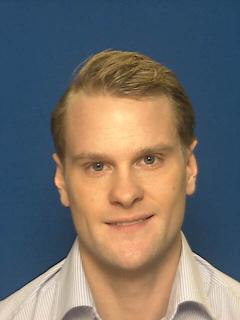}}]{Andreas Ericsson} (M'13) received his M.Sc. degree in engineering physics and his Ph.D. degree in electrical engineering from Lund University, Lund, Sweden, in 2013 and 2017, respectively. He was awarded a student paper award at URSI GASS 2014 in Beijing. He is currently working as a research engineer at TICRA, and his research interests are electromagnetic scattering, antennas, electromagnetic properties of materials, frequency and polarization selective structures.    
\end{IEEEbiography}

\begin{IEEEbiography}[{\includegraphics[width=1in,height=1.25in,clip,keepaspectratio]{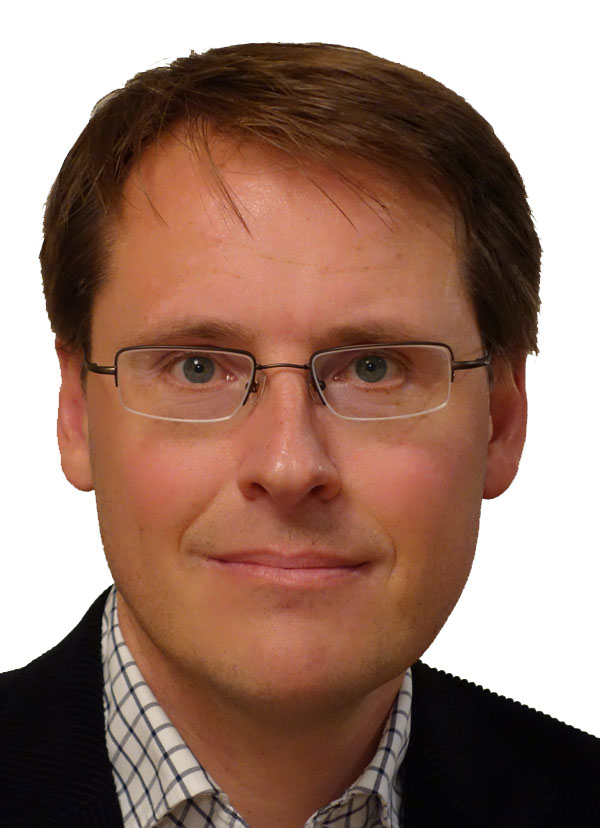}}]{Mats Gustafsson}(SM'17)
received the M.Sc. degree in Engineering Physics 1994, the Ph.D. degree in Electromagnetic Theory 2000, was appointed Docent 2005, and Professor of Electromagnetic Theory 2011, all from Lund University, Sweden. 

He co-founded the company Phase holographic imaging AB in 2004. His research interests are in scattering and antenna theory and inverse scattering and imaging. He has written over 100 peer reviewed journal papers and over 100 conference papers. Prof. Gustafsson received the IEEE Schelkunoff award 2010, the IEEE Uslenghi award 2019, and best paper awards at EuCAP 2007 and 2013. He served as an IEEE AP-S Distinguished Lecturer for 2013-15.
\end{IEEEbiography}

\begin{IEEEbiography}[{\includegraphics[width=1in,height=1.25in,clip,keepaspectratio]{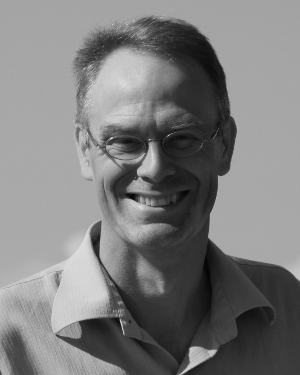}}]{B. L. G. Jonsson} received his Ph.D. degree in electromagnetic theory in 2001 from KTH Royal Institute of Technology, Stockholm, Sweden. He was a postdoctoral fellow at University of Toronto, Canada and a Wissenschaftlicher Mitarbeiter (postdoc) at ETH Zurich, Switzerland. Since 2006 he is with the Electromagnetic Engineering Lab at KTH. He is professor in Electromagnetic fields at KTH since 2015. His research interests include electromagnetic theory in a wide sense, including scattering, antenna theory and nonlinear dynamics. 
\end{IEEEbiography}

\begin{IEEEbiography}[{\includegraphics[width=1in,height=1.25in,clip,keepaspectratio]{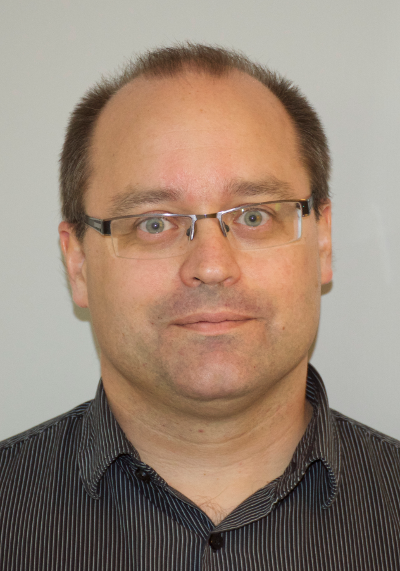}}]{Daniel Sj\"{o}berg} Daniel Sjöberg (M'11, SM'17) received the M.Sc. degree in engineering physics and Ph.D. degree in engineering, electromagnetic theory from Lund University, Lund, Sweden, in 1996 and 2001, respectively. In 2001, he joined the Electromagnetic Theory Group, Lund University, where, in 2005, he became a Docent in electromagnetic theory.
He is currently a Professor and the Head of the Department of Electrical and Information Technology, Lund University. His research interests are in electromagnetic properties of materials, composite materials, homogenization, periodic structures, numerical methods, radar cross section, wave propagation in complex and nonlinear media, and inverse scattering problems.
\end{IEEEbiography}

\end{document}